\documentstyle[twoside,fleqn,espcrc2]{article}

\newcommand{\AmS}{{\protect\the\textfont2
  A\kern-.1667em\lower.5ex\hbox{M}\kern-.125emS}}

\title{New Method for Dynamical Fermions and Chiral-Symmetry Breaking}

\author{V. Azcoiti
\address{Departamento de F\'{\i}sica Te\'orica,
Facultad de Ciencias, Universidad de Zaragoza,\\
        50009 Zaragoza, Spain},
G. Di Carlo$^{\rm b}$, A.F. Grillo
\address{Istituto Nazionale di Fisica Nucleare, Laboratori
Nazionali di Frascati,
        P.O.B. 13 - Frascati (Italy)}
V. Laliena$^{\rm a}$ and X.Q. Luo$^{\rm a}$
        \thanks{Presented by X.Q. Luo. This work has been partly supported
through a INFN(Italy)-CICYT(Spain) collaboration. X.Q. Luo acknowledges
support from the Ministerio de Educacion y Ciencia.}}

\begin{document}

\begin{abstract}
The reasons for the feasibility of the Microcanonical
Fermionic Average ($MFA$)  approach to lattice gauge theory with
dynamical fermions are discussed.  We  then present a new exact
algorithm, which is free from systematic errors and
convergent even in the chiral limit.
\end{abstract}

\maketitle

\section{INTRODUCTION}
The dynamics of a gauge-fermion system on a  $d-$dimensional lattice
is described by its action
\begin{equation}
S=-\beta S_{pl}(U)-N_{f}^{L} \ln \det \Delta(m,U),
\end{equation}
where the fermionic degrees of freedom have been integrated out,
and $N_{f}^L$ is
number of flavors taking into account the species doubling.
$S_{pl}$ is the plaquette energy
[$S_{pl}=\sum_p Re(U_p)/N_c$,
for a compact theory, and
$S_{pl}=-\sum_p \theta_{p}^2/2$,
for
noncompact $QED$].

The fermionic determinant $\det \Delta(m,U)$
accounts for the effects of dynamical fermions; its numerical
evaluation requires in general a huge $CPU$ time
due to nonlocality and dependence on different bare parameters.

Therefore, one of the most challenging tasks
in lattice simulations
is the inclusion of dynamical fermions.
Furthermore, most of the conventional algorithms can not be applied to
the chiral limit ($m=0$) because of the problem of convergence.

In this paper, we discuss two efficient approaches to simulating dynamical
fermions which are convergent even in the chiral limit:
the Microcanonical Fermionic Average ($MFA$) method and
a new exact algorithm.

\section{FEASIBILITY OF THE MFA METHOD}
The main point of the $MFA$ approach \cite{Original} to
dynamical fermions is to compute
the full effective action as a function of
the pure gauge energy $E$ and bare
parameters ($m$, $N_{f}$, and $\beta$):
\begin{eqnarray*}
S_{eff}(E,m,N_f,\beta)= - \ln N(E) -  {d(d-1) \over 2} V \beta E
\end{eqnarray*}
\begin{equation}
+ S^{F}_{eff}(E,m,N_f),
\end{equation}
where
\begin{equation}
N(E)=\int [dU]
\delta(S_{pl} (U) - {d(d-1) \over 2} VE)
\end{equation}
is the density of states at fixed $E$ and
\begin{equation}
S_{eff}^F(E,m,N_f)=
-\ln \langle[\det
\Delta(m, U)]^{N_{f}^{L}}\rangle_E
\end{equation}
is the
effective fermionic action
computed as a microcanonical average over
the probability distribution
\begin{equation}
{\delta(S_{pl} (U) - {d(d-1) \over 2} VE) \over N(E)}.
\end{equation}
Then the partition function becomes a one-dimensional integral
\begin{equation}
Z= \int dE e^{-S_{eff}(E,m,N_f,\beta)}.
\end{equation}
Once the full effective action as a function of $E$ and other bare parameters
is known, the dynamics of the system can be investigated numerically or
analytically. There are several advantages in this approach:

\noindent
(a) $S_{eff}^F$ does not depend on $\beta$;

\noindent
(b) The dependence on $N_{f}^{L}$ is trivial;

\noindent
(c) $S_{eff}^F$ can be simultaneously computed for
any $m$ (and particularly for $m=0$)
if the Lanczos algorithm is employed to evaluate $\det \Delta$.

Now the most important question is how to compute $S_{eff}^F$ defined by
(4). Since the configuration is sampled with the flat distribution (5),
it may require  high statistics
to have the statistical
errors in  $\det \Delta$ under control. A more reliable way
is to perform a cumulant expansion,
which turns out to be an expansion on $N_{f}^{L}$
\begin{eqnarray*}
-S_{eff}^F (E,m,N_f) = N_{f}^{L} \langle \ln \det \Delta\rangle_E
\end{eqnarray*}
\begin{equation}
+ (N_{f}^{L})^{2}
\langle (\ln \det \Delta - \langle \ln \det
\Delta \rangle_{E})^2\rangle_E + ...,
\end{equation}
where the first term is
some kind of mean field approximation,
and the fluctuations correspond to the first corrections.

As is well known, the mean field approximation is a good one if
the interaction is long ranged or the dimensionality is large.
In \cite{PRD}, we have given explicit
physical examples (the local Ising model,
the nonlocal Ising model, the pure gauge, mixed  compact-noncompact model etc.)
to show how the locality and dimensionality play a
role. In the nonlocal Ising model, only the first contribution
to the cumulant expansion exists.

In the most interesting application of
the $MFA$ method, the gauge-fermion system, because of the nonlocality
of $\det \Delta$, it is expected that the cumulant expansion (7) converges
rapidly, which has indeed been observed in all the simulations
\cite{Schwinger,CQED3,PCQED3,NQED3,QCD3,CQED4,NQED4,PCQED4}
of lattice
gauge theory with dynamical fermions ($QED_2$, $QED_3$, $QCD_3$ and $QED_4$).
For instance \cite{PRD},
in noncompact $QED_4$, the first cumulant is dominant, the
second one is 2 $\%$ of the first one, and the third one is compatible
with zero.
Using the $MFA$ method, the results for the Schwinger model \cite{Schwinger}
and for $QED_4$ \cite{CQED4,NQED4,PCQED4}
are in agreement with the exact calculable ones and consistent with
those obtained with the Hybrid Monte Carlo algorithm ($HMC$) respectively.
The simulation of $QCD_4$ is in progress.

\begin{table*}[hbt]
% space before first and after last column: 1.5pc
% space between columns: 3.0pc (twice the above)
\setlength{\tabcolsep}{1.5pc}
% -----------------------------------------------------
% adapted from TeX book, p. 241
\newlength{\digitwidth} \settowidth{\digitwidth}{\rm 0}
\catcode`?=\active \def?{\kern\digitwidth}
% -----------------------------------------------------
\caption{Acceptance rate of the exact algorithm}
\label{tab:AR}
\begin{tabular*}{\textwidth}{@{}l@{\extracolsep{\fill}}rrrr}
\hline
                 & \multicolumn{2}{l}{$S_{eff}^F=0$}
                 & \multicolumn{2}{l}{$S_{eff}^F$ obtained by $MFA$} \\
\cline{2-3} \cline{4-5}
                 & \multicolumn{1}{r}{$m=0$}
                 & \multicolumn{1}{r}{$m=0.02$}
                 & \multicolumn{1}{r}{$m=0$}
                 & \multicolumn{1}{r}{$m=0.02$}         \\
\hline
$N_f=2$, $\beta=0.3$  & $26 \%$ & $36 \%$ & $40 \%$ & $60 \%$ \\
\hline
$N_f=4$, $\beta=0.25$  & $14 \%$ & $15 \%$ & $26 \%$ & $30 \%$ \\
\hline
\end{tabular*}
\end{table*}

\section{EXACT ALGORITHM FOR THE CHIRAL LIMIT}

As mentioned previously, the feasibility of the $MFA$ method
depends on the convergence of the cumulant expansion. In practice,
only the first few cumulants  can  be reliably
calculated and a truncation of the
expansion has to be made. Since a good control of systematic errors is
difficult, especially when $N_f$ is large, it seems interesting to us
to develop some exact algorithm for obtaining quantitative results.

An efficient exact algorithm for dynamical fermions at $m \not= 0$ does
exist, {\it i.e.}
the $HMC$ algorithm. In order to study the chiral properties of
lattice gauge theory without introduction of the external field $m \bar{\psi}
\psi$ or mass extrapolations, it is important to construct an exact algorithm
which is convergent even in the $m \to 0$ limit. To our knowledge, the Lanczos
algorithm is the only efficient technique for evaluating $\det \Delta$
which is convergent in this limit. The problem with this algorithm
is that,
if the canonical simulation is carried out by using the
probability distribution $e^{-(S_g-N_{f}^{L} \ln \det \Delta)}$ for {\it each}
link update then the computer time is prohibitive; if many links are updated
before testing for acceptance,
then its rate is very low.

Here we present a new exact algorithm, based on the following decomposition of
the full action
\begin{equation}
S=S_g+S_{eff}^F+(-N_{f}^{L} \ln \det \Delta-S_{eff}^F),
\end{equation}
where $S_{eff}^F$ is a {\it local} effective action chosen to improve the
acceptance rate of the canonical simulation,
and the partition function can be rewritten as
\begin{equation}
Z =\int [dU] e^{-(S_g+S_{eff}^F)} e^{-(-N_{f}^{L} \ln \det \Delta-S_{eff}^F)}.
\end{equation}
This exact algorithm is implemented as follows.

\noindent
(a) Generate a new configuration $U'$ with the probability distribution
$e^{-(S_g+S_{eff}^F)}$,
which can be done efficiently and is equivalent to a effective pure gauge
theory if we choose $S_{eff}^F$ as that in Sect. 2.

\noindent
(b) Accept $U'$ according to the probability
\begin{equation}
P_A=min \lbrace 1,{e^{N_{f}^{L} \ln \det \Delta(m,U')+S_{eff}^F(m,U')}
\over e^{N_{f}^{L} \ln \det \Delta(m,U)
+S_{eff}^F(m,U)}} \rbrace,
\end{equation}
which is just the Metropolis test.

It can be shown that this approach is free from systematic errors and that
detailed balance is satisfied. Furthermore this algorithm is
convergent even in the chiral limit, since $\det \Delta$ is evaluated by the
Lanczos method.

According to (10), the acceptance rate is higher if  $S_{eff}^F$ is a
good {\it local} approximation to the fermionic action.
Here we choose $S_{eff}^F$ as that in the $MFA$ method discussed in Sect. 2.

We have tested these ideas in noncompact $QED_3$ using this new
exact algorithm. To avoid huge changes in $\ln \det \Delta$, we change only
part of the links in each iteration (a). The results for the $10^3$ lattice
(only 1000 links are systematically changed when $U'$ is generated)
are the following:

\noindent
(i.) For quenched ($N_f=0$) and $N_f=2$ theories,
the data for the chiral condensate $\langle \bar{\psi} {\psi}\rangle$ at
nonzero
mass  agree with those \cite{Kogut} obtained by the hybrid algorithm.

\noindent
(ii.) The results for $N_f=4$ are slightly different from those
in \cite{Kogut}. This is not surprising because systematic errors increase
as $O(N_{f}^2)$ in the hybrid algorithm.

\noindent
(iii.) In the canonical simulations,
we find that the choice of $S_{eff}^F$ plays
an important role in the improvement of the acceptance rate.
Table~\ref{tab:AR} shows
the results for the acceptance rate for different
bare parameters. The data for direct canonical simulations
($S_{eff}^F$=0) are also included in this Table. As one sees, the use of the
effective fermionic action obtained by the $MFA$ method gives much better
results for this quantity (it increases by about $40 \%$).

\section{CONCLUSIONS}
To summarize, we have argued that one of the most important reasons
for the feasibility of the $MFA$ approach is the nonlocality of $\det \Delta$.
Based on the $MFA$ method, a new exact algorithm, which is
convergent even at $m =0$ is presented and possible ways
of increasing the acceptance rate are discussed. An interesting application
of this exact algorithm, the investigation of the vacuum chiral
properties directly at $m=0$ by calculating the probability distribution
function of the chiral condensate, can be found in \cite{PDF}.

\end{document}